\documentclass[twocolumn,showpacs,preprintnumbers,amsmath,amssymb]{revtex4}

\usepackage{graphicx}
\usepackage{dcolumn}
\usepackage{bm}
\usepackage[usenames]{color}

\begin{document}


\title{Interplay between charge-order, ferroelectricity and ferroelasticity: \\
tungsten bronze structures as a playground for multiferroicity}

\author{Kunihiko Yamauchi}
\author{Silvia Picozzi}%
\affiliation{%
Consiglio Nazionale delle Ricerche - Istituto Nazionale di Fisica della Materia (CNR-INFM), CASTI Regional Lab., 67100 L'Aquila, Italy\\
}%

\date{\today}

\newcommand{\banana}{Ba$_{2}$NaNb$_{5}$O$_{15}$}
\newcommand{\KFeF}{K$_{0.6}$Fe$_{0.6}^{\rm II}$Fe$_{0.4}^{\rm III}$F$_{3}$}

\begin{abstract}
Large electron-electron Coulomb-interactions in correlated systems can lead to a periodic arrangement of localized electrons,
the so called
``{\em charge-order}". The latter
 is here proposed as a driving force behind ferroelectricity in iron fluoride \KFeF. By means of density functional theory, we propose different non-centrosymmetric $d^{5}$/$d^{6}$ charge-ordering patterns, each giving rise to polarization along different crystallographic axes and with different magnitudes. 
 Accordingly, we introduce the concept of ``ferroelectric anistropy" (peculiar to improper ferroelectrics
 with polarization induced by electronic degrees of freedom), denoting the small energy difference  between competing  charge-ordered states that might be stabilized upon electrical field-cooling. Moreover, we suggest a novel type of {\em charge-order-induced ferroelasticity}: first-principles simulations predict a monoclinic distortion to be driven by  a specific charge-ordering pattern, which, in turn, unambiguously determines the direction of ferroelectric polarization. \KFeF\  therefore emerges as a prototypical compound, in which the intimately coupled electronic and structural degrees of freedom result in a manifest and peculiar multiferroicity.

\end{abstract}

\pacs{Valid PACS appear here}
\maketitle
Materials which combine magnetism and ferroelectricity, belonging to the intriguing class of multiferroics, can be classified into two categories:\cite{reviewclaude} ``Structural magnetic ferroelectrics" (SMF), 
where the primary order parameter in the ferroelectric (FE) phase transition is 
related to a structural instability (which can be polar or non-polar, see for example
 the prototypical BiFeO$_3$ or
the ``geometric FE" YMnO$_3$), and ``electronic magnetic ferroelectrics" (EMFs),
where the primary order parameter is related to electronic
degrees of freedom, such as spin, charge or orbital order. \cite{maxim,kimura,efremov,prlslv}

Whereas plenty of studies have been recently performed in the field of spin-driven ferroelectricity, 
ferroelectricity induced  by charge-order (CO) still constitutes a largely unexplored territory. Even in the two paradigmatic
cases in which the pattern of Fe$^{2+}$ and Fe$^{3+}$ ions  was suggested to break space inversion symmetry, 
e.g., LuFe$_{2}$O$_{4}$\cite{ikeda} and
Fe$_{3}$O$_{4}$\cite{alexe,yamauchi}, the actual occurrence of ferroelectricity seems controversial.

Within this context, non-oxides - and fluorides in particular - are interesting candidates as potentially novel
improper multiferroics. Indeed,  \KFeF\, which
crystallizes in a non-centrosymmetric tetragonal tungsten bronze (TTB) structure, has been reported to show CO, although the exact CO pattern is still debated.\cite{Mezzadri} The family of TTB compounds, obtained by substituting either tungsten with transition metals or oxygen with other anions, exhibit a lot of functionalities, such as ferroelectricity, ferroelasticity, pyro-/piezo-electric properties\cite{Jamieson}. 
Despite such a remarkable technological appeal, the electronic structure of TTB compounds has not been deeply investigated from the theory point of view, mainly because of the complex crystal structure. 
Guided by the high potential of TTB materials in the multiferroics field, 
 we here focus on \KFeF\, showing ferroelasticity and ferroelectricity driven by non-centroymmetric Fe-$d^{5}$/$d^{6}$ CO patterns. To outline the novel physics in \KFeF\, we will first discuss  the prototypical TTB compound, \banana, where ferroelectricity is driven by conventional Nb$^{5+}$ off-centering and where no coupling between ferroelectricity and ferroelasticity is observed.

\begin{figure}
\centerline{\includegraphics[width=0.6\columnwidth]{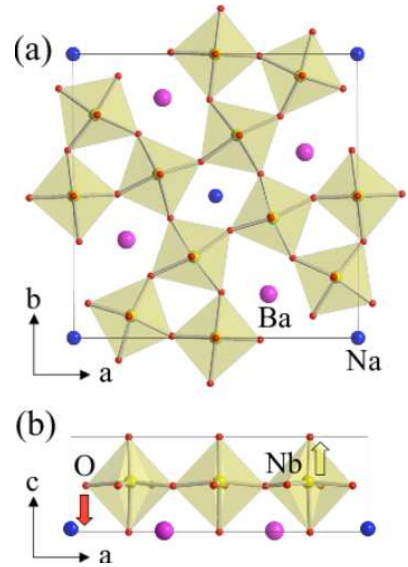}}
\caption{\label{figbananastruc} 
C-centered primitive unit cell of the optimized crystal structure of \banana\ in (a) $ab$ plane and in (b) $ac$ plane (only three Nb-O$_{6}$ octahedrons are shown). Upon ionic relaxation, all Nb (O) ions are displaced toward +z (-z) direction, so as to induce a large $P_{z}$.
}
\end{figure}

{\em Methodology and structural details - } DFT simulations were performed using the VASP code \cite{vasp} and the PAW pseudopotentials \cite{paw} within the GGA+$U$ formalism\cite{ldau} ($U$=5 eV and $J$=0 eV for Fe $d$-states).   
The cut-off energy for the plane-wave expansion  of the wave-functions was set to 400 eV and a {\bf k}-point shell of (2, 2, 4) was used for the Brillouin zone integration.  
The internal atomic coordinates were fully optimized until the atomic forces were less than 0.01 eV/\AA\ while the lattice parameters were taken from experiments.   
The FE polarization $P$ was calculated using the Berry phase method\cite{berry}, by comparing the FE and paraelectric (PE) state; the latter is constructed by imposing the $x,  y, z$ reflection in the atomic structure. For simplicity, 
 \KFeF\  was treated as a ferromagnet; further complexity of the experimentally suggested
ferri- or weak ferro-magnetic spin configuration is not expected to affect CO (and 
related ferroelectricity), as in LuFe$_{2}$O$_{4}$.\cite{xiang_lufe2o4}

{\em $d^{0}$-ness at \banana\ - } At room temperature, \banana (BNN)\cite{scott_banana} crystallizes in the polar orthorhombic $Cmm2$
 structure with $a=17.626$, $b=17.592$, $c=3.995$ \AA \cite{barns}. 
At low temperature, BNN shows quasi-commensurate and incommensurate phases with {\em ferroelastic} transition, leading to a
crystal cell in the $ab$ plane larger than the above mentioned unit cell.\cite{burgeat} 
 Since ferroelectricity was reported not to be coupled with ferroelasticity\cite{toledano} and since our main 
 focus is on FE properties, 
we optimized BNN  in the $Cmm2$
structure. 
As shown in Fig.\ref{figbananastruc},
 the optimized NbO$_{6}$ octahedrons are significantly distorted from the tetragonal symmetry. 
The polar distortion along the $z$ axis is driven by the off-center shift of Nb atoms, due to a strong hybridization between Nb empty $4d$ and O $2p$ states, with an average Nb$^{5+}$ ionic displacement (with respect to the side O ions) of 0.16 \AA. The FE behaviour can be clearly interpreted on the basis 
of the ``$d^{0}$-ness'' criterion\cite{nicola_d0ness}. The latter doesn't only cause electric polarization, but also results in a wide energy gap:  2.7eV in the FE phase, with respect to  2.3eV in the PE state. 
The calculated polarization is 
34.3$\mu$C/cm$^2$ 
along the $c$ direction; to our best knowledge, this is the first theoretical estimate reported in the literature and is in good agreement with the experimental value of 40$\mu$C/cm$^2$\cite{sambasiva}.

{\em Charge order at \KFeF\ - }
\begin{figure}
\centerline{\includegraphics[width=1\columnwidth]{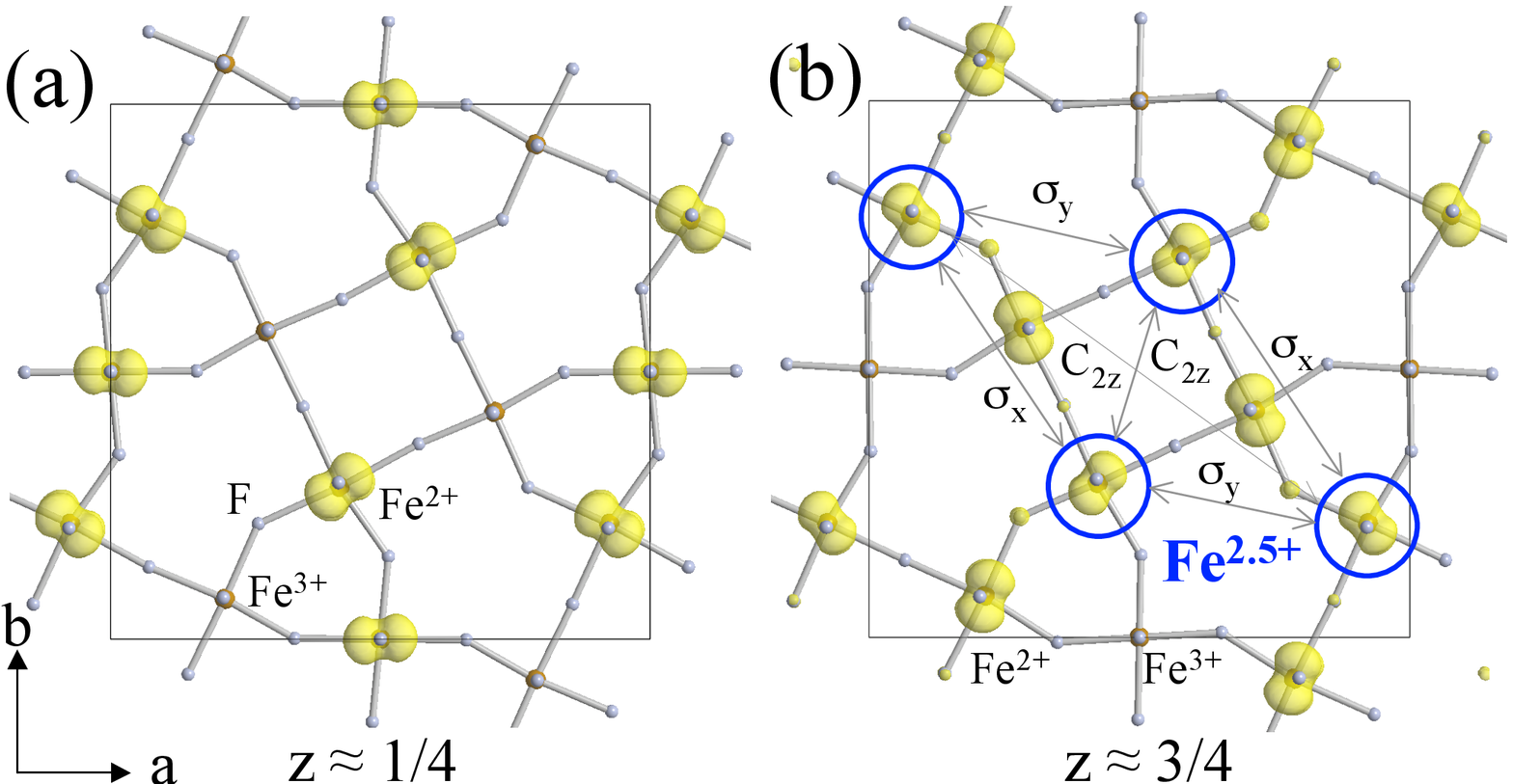}}
\vspace{.4cm}
\centerline{\includegraphics[width=1\columnwidth]{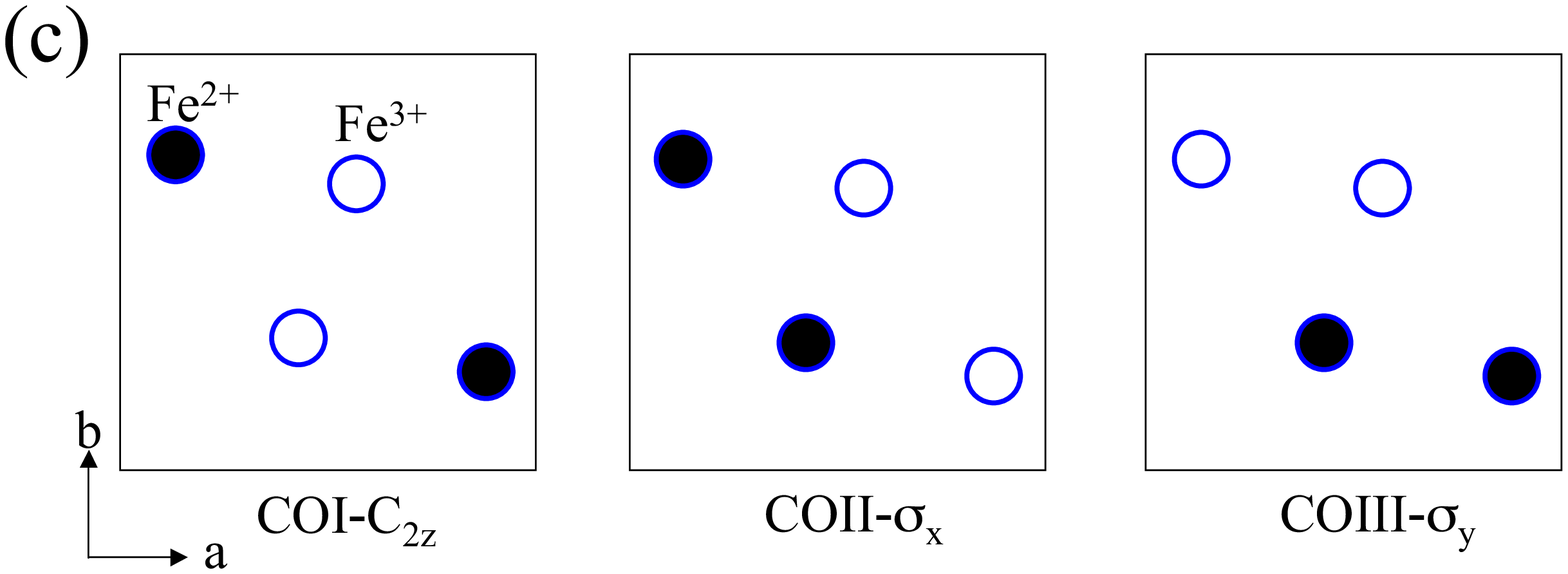}}
\caption{\label{figexpCO} 
Down-spin charge of Fe $t_{2g}$ state, 1 eV below $E_{\rm F}$, for the $Pba2$ symmetry 
at (a) $z\approx1/4$ and (b) $z\approx3/4$ planes.  
Circled ions  are mixed-valence Fe$^{2.5+}$ ions. The symmetry operations which relate the four Fe$^{2.5+}$ are also shown.
(c) Full CO patterns 
starting from partial CO shown in (b) can be obtained by turning 
four Fe$^{2.5+}$ sites into two Fe$^{2+}$ and two Fe$^{3+}$ sites at $z\approx3/4$ plane. The positions of (blue) circles correspond to the one  in (b). 
}
\end{figure}
In iron fluoride, the complexity of the different phase transitions is still a matter of debate. 
Earlier experiments  showed coupled ferroelectric/ferroelastic transitions to occur at 490 K,\cite{ravez} along with CO.\cite{calage} More recently, several transitions were reported:\cite{Mezzadri}
 a first structural transition around 570K from tetragonal to orthorhombic, a second transition at 490K where Fe$^{2+}$/Fe$^{3+}$ CO occurs, and a third transition around 290K to monoclinic structure, coupled with ferroelasticity.    
Experimentally, the orthorhombic $Pba2$ structure was reported, with $a=12.751$, $b=12.660$ and $c=7.975$\AA\cite{Mezzadri}, the unit cell including two TTB layers along $c$. 
From the symmetry point of view, the $Pba2$ group has four symmetry operations: \{$E, C_{2z}, \sigma_{x}+(\frac{1}{2} \frac{1}{2} 0), \sigma_{y}+(\frac{1}{2} \frac{1}{2} 0)$\} so that  FE polarization is in principle allowed to be induced only along the $z$ axis.
Given the $d^5$/$d^6$ electronic configuration of Fe ions,  $P_{z}$ by ``$d^{0}$-like" hybridization is of course not expected here, so that alternative mechanisms, such as CO, should be invoked to explain ferroelectricity.
As discussed in ref.\cite{Mezzadri}, the Fe ions show CO in two FeO layers. However, the CO is not ``full"  in the experimental $Pba2$ structure:
although 12 Fe ions are supposed to be Fe$^{2+}$($d^{6}$) and 8 Fe ions to be Fe$^{3+}$($d^{5}$) due to the stoichiometry (K$_{12}$Fe$_{12}^{\rm II}$Fe$_{8}^{\rm III}$F$_{60}$/cell),  
the $Pba2$ crystal structure gives 10 Fe$^{2+}$, 6 Fe$^{3+}$ and 4 Fe$^{2.5+}$ ions (cfr
 Fig.\ref{figexpCO}(a),(b)). 
As expected, the mixed-valence Fe ions lead to  a metallic  behaviour  (not shown), kept even after the structural optimization.
In order to obtain an insulating state (obviously needed for ferroelectricity to develop), the symmetry must be reduced so as to get   a fully charge-ordered Fe sublattice. 
The existence of a larger supercell which may stabilize a complete CO 
was previously suggested;\cite{Mezzadri}   however, this would require an extremely high and unaffordable computational load. Therefore,
we limited our study to the $Pba2$
unit cell, artificially breaking  symmetries so as to obtain a ``full" CO. 
Here we assumed three CO patterns, CO-I, CO-II and CO-III, each keeping one of the three symmetry operations and breaking the other two  among $C_{2z}, \sigma_{x}+(\frac{1}{2} \frac{1}{2} 0)$ and $\sigma_{y}+(\frac{1}{2} \frac{1}{2} 0)$, respectively (cfr Fig.2(c)). 
Whereas the induced polarization $P_{z}$ is allowed by the prototype $Pba2$ crystal symmetry, 
$P_{y}$ and $P_{x}$ are additionally allowed in COII and COIII, respectively. 

\begin{table}[h] \begin{center}
\caption{Total energy difference (meV/Fe), energy gap (eV) and induced FE polarization calculated by Berry phase $P^{\rm Berry}$ and by dipoles $P^{\rm dipole}$ ($\mu$C/cm$^{2}$) at partial CO pattern in experimental crystal structure, optimized structure keeping experimental symmetry, full CO patterns; CO-I, CO-II, CO-III (with given symmetry). } 
\label{tblen}
\begin{tabular}{cccccc}
\hline 
                                  & pCO$_{\rm exp}$ & $_{\rm opt}$ & COI-$C_{2z}$ & COII-$\sigma_{x}$ & COIII-$\sigma_{y}$\\
	         			 & $Pba2$& $Pba2$ &  $P2$ &  $Pc$ & $Pc$           \\                    
$\Delta E^{\rm tot}$    & 0 & -31.3 & -44.3 & -57.3 & -51.6 \\
$E^{\rm gap}$             & 0 & 0      & 0.91 & 1.28 & 1.08 \\ 
$P^{\rm Berry}$          &  ---  &  ---  & (0 0 0.09) &(0 -0.50 -0.19) &(-5.14 0 0.03) \\
$P^{\rm dipole}$          & ---  & --- & --- & $P_{y}=$-0.57 & $P_{x}=-5.43$ \\
\hline 
\end{tabular}
\end{center}
\end{table}

After ionic optimization at each CO-pattern, the charge separation between Fe$^{2+}$ and Fe$^{3+}$, calculated by integrating the charge density in 1\AA\ atomic radius, is  equal to 0.365 $e^-$; this is rather large compared to the value of Fe$_{3}$O$_{4}$ ($\approx$0.2 $e^-$) \cite{yamauchi}. Indeed, this is consistent with  the expected weaker Fe-F hybridization compared to Fe-O and put forward fluorides as better candidates for CO - compared to oxides - where a larger charge-disproportionation can be achieved.  

\begin{figure}
\centerline{\includegraphics[width=0.9\columnwidth]{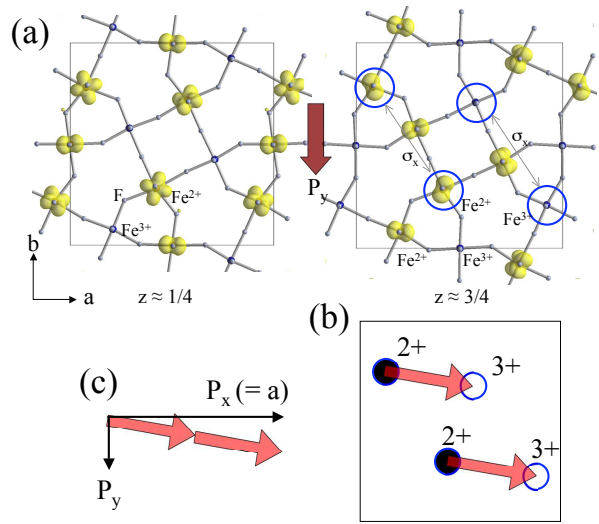}}
\vspace{.1cm}
\caption{\label{figCOSigmax} 
(a) Down-spin charge of Fe $t_{2g}$ states, 1 eV below $E_{\rm F}$ in the optimized structure with the ground-state COII-$\sigma_{x}$ pattern. 
Circles show the Fe sites where Fe$^{2.5+}$ mixed-valence ions are originally located in the experimental structure. 
(b) Electric dipoles arising among circled Fe sites, due to the different valence occurring upon full CO. 
(c) The polarization $P_{x}$ has the same size as the $a$ Bravais vector with 1 electron charge ($2\pi$ in Berry phase), so that $P_{x}$ is zero; on the other hand, the net $P_{y}$ is non-zero. 
}
\end{figure}

{\em CO-induced ferroelectricity - } We found that, among the assumed CO patterns as well as  the experimental/optimized state, 
COII is the most stable state  with the largest energy gap (cfr Tab. \ref{tblen}). 
Besides, COII shows a sizeable polarization $P_{y}^{\rm Berry}=-0.50$ $\mu$C/cm$^{2}$. 
This can be roughly understood as induced by local electric ``point-charge" dipoles which connect Fe$^{2+}$ and Fe$^{3+}$ ions,  as previously suggested for Fe$_{3}$O$_{4}$\cite{yamauchi}. The electric dipoles are calculated considering only four Fe ions, {\em i.e.} those  originally located on the mixed valence sites. In fact, the CO pattern of the other Fe ions are identical to the high-symmetry $Pba2$ CO pattern, so that their net contribution to {\bf P} would cancel out.
As shown in Fig.\ref{figCOSigmax}(b) and (c), $P_{y}^{\rm dipole}$ is calculated as
 -0.57 $\mu$C/cm$^{2}$ in good agreement with $P^{\rm Berry}$: this shows that polarization has a purely electronic origin and that ionic displacements play a minor role.
 \color{black}
Note that the sign of $P_{y}$ can be switched by exchanging Fe$^{2+}$ and Fe$^{3+}$ ions;  under an applied electric field $E_{y}$, charge shifts among these four Fe sites is indeed expected to occur. 
In an analogous way, we calculated $P_{x}$ at CO-III pattern as $P_{x}^{\rm dipole}\simeq P_{x}^{\rm Berry}\simeq -5\mu$C/cm$^{2}$, a  value much larger than $P$ at CO-II. In this case, too, the agreement between $P_{x}^{\rm dipole}$ and $P_x^{\rm Berry}$ is remarkable. Finally, we note that, as expected, the CO-induced polarization, though sizeable, is  much smaller than in \banana\ where the ionic degrees of freedom are the source of ferroelectricity.

\begin{figure}
\vspace{.5cm}
\centerline{\includegraphics[width=0.8\columnwidth]{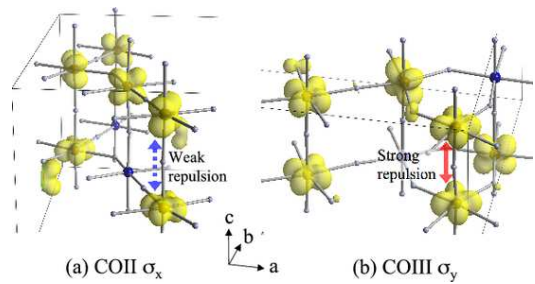}}
\vspace{.1cm}
\caption{\label{3Dview} 
Perspective view of OO of Fe-$t_{2g}$ down-spin states (within an energy range up to 1 eV below $E_{\rm F}$) at (a) COII and (b) COIII. The strong/weak next-neighbor orbital overlap and the consequent intersite Coulomb repulsion is highlighted.}
\end{figure}

{\em Charge and Orbital Order - } The CO arrangement  also causes a change in the Fe-$t_{2g}^{1}$ orbital ordering (OO). 
When comparing Fig.\ref{figexpCO} and Fig.\ref{figCOSigmax}, 
the occupied $t_{2g}$ orbitals (both at $z$=1/4 and $z$=3/4 planes) modify their shape/direction to avoid as much as possible the nearest $t_{2g}$-$t_{2g}$ overlap. 
Figure \ref{3Dview} shows the three dimensional network of Fe-$t_{2g}$  OO. 
It is clear that the $t_{2g}$-orbital relative to the ``extra charge'' on top of Fe$^{2+}$ site at COII pattern is aligned along a diagonal direction in the octahedron, so as to avoid the strong inter-site Coulomb interaction 
with neighboring $t_{2g}$-orbitals at Fe$^{2+}$ sites. 
Therefore, we argue that the COII pattern is energetically lower than COIII, due to minimization of  Coulomb repulsion. 

{\em CO-induced Ferroelasticity - } From the above discussions, we conclude that the ground state of \KFeF\ is FE with a polar COII pattern (although this has yet to be confirmed experimentally). 
Furthermore, the polar CO patterns induce  ``unbalanced"  (from the valence point of view) planes at $z=1/4$ and at $z=3/4$, so that a monoclinic crystal distortion is expected (in this case along the direction of $P$, which is determined by the COII pattern as well). This is proposed as responsible for the reported ferroelastic phase\cite{Mezzadri}. 
To check this effect, we optimized the $a, b, c$ lattice vectors under CO-I and CO-II patterns, keeping the cell volume fixed to the experimental value.
It turns out that the optimized lattice is in fact monoclinically distorted, with an angle $\angle bc$=90-0.059$^{\circ}$ at CO-II, and an $\angle ac$=90-0.040$^{\circ}$  at CO-III. 
This reveals a significant fact: {\em ferroelasticity is strongly coupled with CO, which in turn determines the direction of the FE polarization}. 
We note that the ferroelastic distortion in the $ac$ or $bc$ plane is peculiar to this CO system and distinct from the conventional monoclinic distortion in the $ab$ plane, the latter occurring in other ferroelastic TTB compounds, $e. g.$ \banana, as well. Although these monoclinic  distortions in \KFeF were not experimentally observed, we remark that they are rather small and possibly below the detection limit.

{\em Ferroelectric anisotropy - } The proposed polar COII pattern may not be spontaneously long-ranged; however, 
it may be possible to stabilize large FE domains by  field-cooling upon applying $E_{y}$, in analogy with
LuFe$_{2}$O$_{4}$ (where
 the FE and the anti-FE states are energetically close and the stabilization of the FE phase occurs via field-cooling). 
We further speculate that the energetically competing CO-III pattern inducing a large $P_{x}$ might be realized by applying $E_{x}$. 
In the case of a (strong enough) applied electric field  rotating  in the $xy$ plane, 
the induced $P$ along the field  would therefore change its saturation value. 
This ``ferroelectric anisotropy'' (FEA),  denoted as the energy required to modify the  direction (as well as size) of the permanent polarization by switching the crystal between different CO phases (in this case COII and COIII), 
may found applications in future devices, such as multiple-state memories where the information can be stored by exploiting not only  the sign of $P$, but also its direction. The FEA is peculiar for  ``improper" ferroelectricity  induced by electronic degrees of freedom; in analogy with CO-induced polarization, the possibility to control the direction of {\bf P} in spin-spirals manganites by means of a magnetic field was already proven and suggestions towards devices harnessing FEA already came.\cite{argiriou}

In summary, tungsten-bronze systems, previously known for hybridization-driven $d^0$ ferroelectricity occurring in \banana, branch into the class of improper multiferroics. 
Indeed, we have put forward CO as the origin of  ferroelectricity (as well as of some ferroelastic modes) in a non-oxide TTB compound, \KFeF. Several energetically-competing CO patterns were predicted from first-principles, with potentially different directions and magnitudes of the polarization. This
``ferroelectric anisotropy", typical of electronic magnetic ferroelectrics, shows a high technological appeal in multiple-state devices. In addition, the strong interplay between CO, ferroelasticity and ferroelectricity in \KFeF\ makes it an excellent compound where multiferroic effects are manifestedly at play.


We thank F. Mezzadri, E. Gilioli and G. Calestani for useful discussions.
The research leading to these results has received funding from the European Research Council under the EU 7$^{th}$ Framework Programme  (FP7/2007-2013) / ERC grant agreement n. 203523.
Computational support from Caspur  Supercomputing Center  (Rome) and Cineca Supercomputing Center (Bologna) is acknowledged. 




\begin{thebibliography}{200}
\bibitem{reviewclaude} S. Picozzi and C. Ederer,  J. Phys.: Cond. Mat. {\bf 21}, 303201 (2009).

\bibitem{maxim} M. Mostovoy  and S.W. Cheong  {\em Nature Mater} {\bf 6}, 13 (2007).

\bibitem{kimura} T. Kimura, T. Goto, H. Shintani, K. Ishizaka, T. Arima, and Y. Tokura, Nature {\bf 426}, 55 (2003).

\bibitem{efremov} D.V. Efremov, J. van der Brink, D. I. Khomskii, Nat. Mater. {\bf 3}, 853 (2004).

\bibitem{prlslv} S. Picozzi, I. A. Sergienko, K. Yamauchi, B. Sanyal and E. Dagotto, Phys. Rev. Lett. {\bf 77}, 227201 (2007).

\bibitem{ikeda}  N. Ikeda, {\em et al.}
Nature (London) {\bf 436}, 1136 (2005).

\bibitem{alexe}  M. Alexe, M. Ziese, D. Hesse, P. Esquinazi, K. Yamauchi, T. Fukushima, S. Picozzi, and U. G\"osele, Adv. Mater. (to be published).

\bibitem{yamauchi} K. Yamauchi, T. Fukushima, S. Picozzi, Phys. Rev. B {\bf 79}, 212404 (2009).

\bibitem{Mezzadri} F. Mezzadri, S. Fabbrici, E. Montanari, L. Righi, G. Calestani, E. Gilioli, F. Bolzoni and A. Migliori, Phys. Rev. {\bf B 78}, 064111 (2008).

\bibitem{Jamieson} P. B. Jamieson, S. C. Abrahms, and J. L. Bernstein, J. Chem. Phys. {\bf 48}, 5048 (1968). 

\bibitem{vasp} G.Kresse and J.Furthm\"uller, Phys.Rev.B {\bf 54}, 11169 (1996).

\bibitem{paw} P. E. Bl\"ochl, Phys. Rev. B {\bf 50}, 17953  (1994).

\bibitem{ldau} V.I. Anisimov, F. Aryasetiawan  and A.I. Lichtenstein,  J. Phys.: Cond. Mat. {\bf 9}, 767 (1997).

\bibitem{berry} R.D. King-Smith and D. Vanderbilt, Phys. Rev. B {\bf 47}, 1651 (1993).

\bibitem{xiang_lufe2o4} H. J. Xiang and M.-H. Whangbo, Phys. Rev. Lett. {\bf 98}, 246403 (2007).

\bibitem{scott_banana} J. F. Scott, J. Phys.: Condens. Matter {\bf 20} 021001 (2008).

\bibitem{barns} R. L. Barns, J. Appl. Cryst.  {\bf 1}, 290 (1968).

\bibitem{burgeat} J. Burgeat and J. C. Toledano, Solid State Comm. {\bf 20}, 281 (1976) . 

\bibitem{toledano} J. C. Toledano, Phys. Rev. B {\bf 12}, 943 (1975).

\bibitem{nicola_d0ness} N. A. Hill, J. Phys. Chem. B, {\bf104}, 6694 (2000).

\bibitem{sambasiva} K. Sambasiva Rao, K. Hyun Yoon, J. Mater. Sci., {\bf 38}, 391 (2003).

\bibitem{ravez} J. Ravez, S.C. Abrahams, R. de Pape, J. Appl. Phys. {\bf 65}, 3987 (1989).

\bibitem{calage} Y. Calage, {\em et al.}
J. Appl. Phys. {\bf 67}, 430 (1990).

\bibitem{argiriou} E. Schierle, {\em et al.},
http://arXiv:0910.5663



 
 















\end{thebibliography}
\end{document}